\documentclass[twocolumn]{aastex631}

\usepackage{xcolor}
\usepackage{amsfonts,amssymb,amsmath}
\usepackage{graphicx}

\newcommand{\result}[1]{#1}

\newcommand{\externalresult}[1]{#1}

\begin{document}

\title{
Exotic Stable Branches with Efficient TOV Sequences
}

\author{Reed Essick}
\email[E-mail:~]{essick@cita.utoronto.ca}
\affiliation{Canadian Institute for Theoretical Astrophysics, 60 St. George St, Toronto, Ontario M5S 3H8}
\affiliation{Department of Physics, University of Toronto, 60 St. George Street, Toronto, ON M5S 1A7}
\affiliation{David A. Dunlap Department of Astronomy, University of Toronto, 50 St. George Street, Toronto, ON M5S 3H4}

\begin{abstract}
    Modern inference schemes for the neutron star equation of state (EoS) require large numbers of stellar models constructed with different EoS, and these stellar models must capture all the behavior of stable stars.
    I introduce termination conditions for sequences of stellar models for cold, non-rotating neutron stars that can identify all stable stellar configurations up to arbitrarily large central pressures along with an efficient algorithm to build accurate interpolators for macroscopic properties.
    I explore the behavior of stars with both high- and low-central pressures.
    Interestingly, I find that EoS with monotonically increasing sound-speed can produce multiple stable branches (twin stars) and that large phase transitions at high densities can produce stable branches at nearly any mass scale, including sub-solar masses, while still supporting stars with $M > 2\,M_\odot$.
    I conclude with some speculation about the astrophysical implications of this behavior.
\end{abstract}



\section{Introduction}
\label{sec:introduction}

Neutron stars (NSs) are extremely dense stellar remnants typically observed with masses between 1--2 $M_\odot$ with radii $\sim 12\,\mathrm{km}$~\citep{Legred:2021}.
As such, NSs reach characteristic densities\footnote{I write expressions in units where $G = c = 1$ but provide explicit units when relevant.} \result{$\left<\rho\right> = 3M/4\pi R^3 \approx 5\cdot10^{14}\,\mathrm{g}/\mathrm{cm}^3$}, roughly twice the density of atomic nuclei (nuclear saturation density: $n_\mathrm{sat} \approx 0.16\,\mathrm{fm}^{-3}$ or $\rho_\mathrm{sat} = m_n n_\mathrm{sat} \approx 2.8\times10^{14}\,\mathrm{g}/\mathrm{cm}^3$ where $m_n$ is the nucleon rest-mass).
Therefore, many-body nuclear interactions play a key role in NS structure.
At the same time, NSs can reach compactness \result{$C = M/R \sim 1/4$}, which is close to the upper limit expected from a Schwarzschild Black Hole (BH): $C_\mathrm{Sch} = 1/2$.
Additionally, the speed of sound ($c_s$) within the NS core almost certainly exceeds $1/\sqrt{3}$~\citep{Legred:2021} and therefore reaches an appreciable fraction of the speed of light.
Clearly, then, relativistic effects should also be significant.

\begin{figure*}
    \includegraphics[width=1.0\textwidth, clip=True, trim=0.0cm 1.1cm 0.0cm 0.0cm]{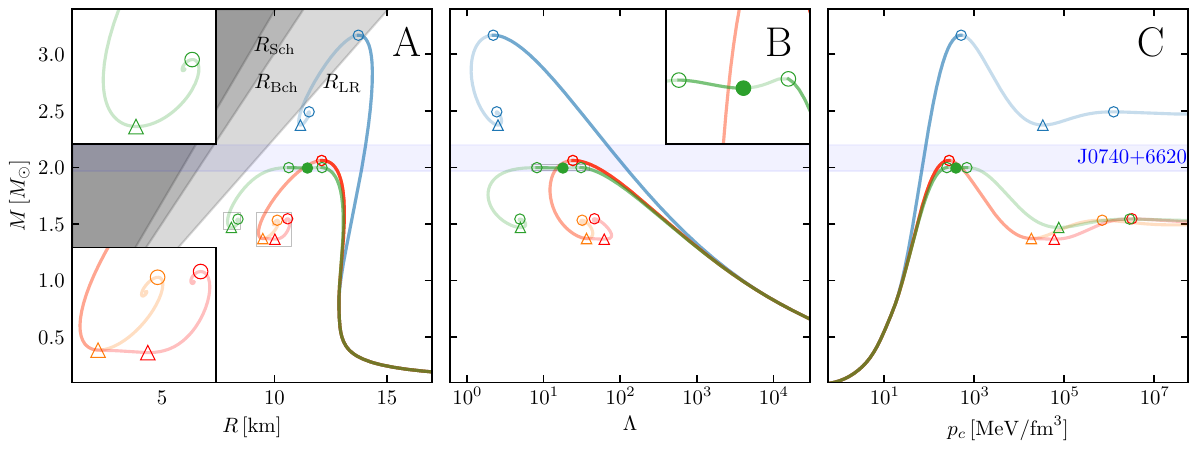}
    \includegraphics[width=1.0\textwidth]{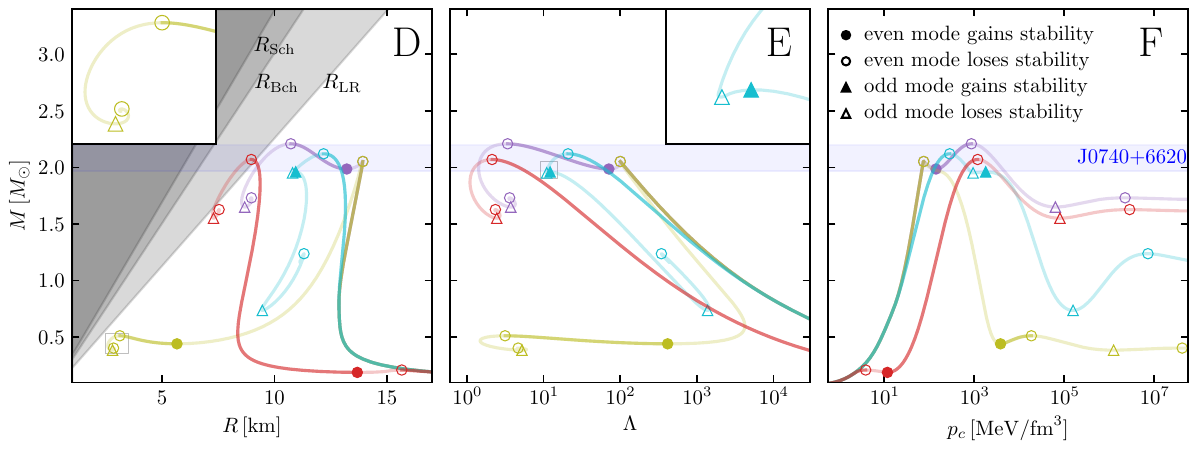}
    \caption{
        EoS based on chiral effective field theory~\citep[$\chi$EFT; ][]{Keller:2023} below $p = 2.2\,\mathrm{MeV}/\mathrm{fm}^3$ 
        and a BPS crust~\citep{Baym:1971}
        (\emph{top}) without and (\emph{bottom}) with 1$^\mathrm{st}$-order phase transitions.
        I show (\emph{left}) $M$-$R$, (\emph{middle}) $M$-$\Lambda$ (tidal deformability), and (\emph{right}) $M$-$p_c$ curves.
        Stable branches are dark solid lines and unstable branches are light lines of the same color.
        The points where stability changes are: (\emph{circles}) even modes ($\mathrm{n}=0, 2, ...$) and (\emph{triangles}) odd modes ($\mathrm{n}=1, 3, ...$) with (\emph{filled symbols}) where stability is regained and (\emph{empty symbols}) where stability is lost.
        For clarity, I only show modes with $\mathrm{n} \leq 2$.
        I also show the Schwarzschild ($R_\mathrm{Sch} = 2M$), Buchdahl ($R_\mathrm{Bch} = 9M/4$), and light ring ($R_\mathrm{LR} = 3M$) radii along with the approximate 90\% credible region for the mass of J0740+6620~\citep{Cromartie:2019, Fonseca:2021}.
    }
    \label{fig:mass-radius}
\end{figure*}

The structure of cold, non-rotating NSs is governed by the Tolman-Oppenheimer-Volkoff (TOV) equations~\citep{Oppenheimer:1939, Tolman:1939}\footnote{It is common practice to solve the TOV equations in a slightly different form, sometimes called the (log)enthalpy formulation~\citep{Lindblom:2014}. See also~\citet{Kastaun:2024} for a recent review of numerical techniques. I find that the enthalpy formulation provides \result{a factor of $\sim$ 5-10 speedup} compared to the ``standard'' formulation described above.}
\externalresult{
\begin{align}
    \frac{dm}{dr} & = 4\pi r^2 \varepsilon \\
    \frac{dp}{dr} & = - \frac{m \varepsilon}{r^2} \left(1 + \frac{p}{\varepsilon}\right)\left(1 + \frac{4\pi r^3 p}{m}\right)\left(1 - \frac{2m}{r} \right)^{-1}
\end{align}
}
where $p$ is the pressure, $\varepsilon$ the energy density, $r$ the radial coordinate, and $m$ the enclosed gravitational mass (distinct from the enclosed rest-mass as it includes the gravitational binding energy).
Because NSs are believed to be supported primarily by degeneracy pressure, these equations are closed with a barytropic equation of state (EoS) that specifies $p$ as a function of $\varepsilon$.\footnote{The EoS can also be specified in terms of other thermodynamic variables, like the number density ($n$), the chemical potential ($\mu = d\varepsilon/dn = (\varepsilon+p)/n$), or the sound-speed ($c_s^2 = dp/d\varepsilon$).}
Finally, the TOV equations are solved with boundary values in the center and at the surface of the star:
\externalresult{
\begin{align}
    \lim\limits_{r\rightarrow 0^+} \frac{dp}{dr} & = 0 \quad \text{(center)} \\
    \lim\limits_{r\rightarrow R^-} p & = 0 \quad \text{(surface)}
\end{align}
}
The only remaining free parameter is the central pressure ($p_c$), and one can construct relationships between, e.g., the radius ($R$) and gravitational mass ($M = m(R)$) by repeatedly solving for a sequence of $p_c$.
One is often instructed to continue this process until a local maximum in $M$ is observed, which corresponds to a loss of stability~\citep{Bardeen:1966, Harrison:1965, Sorkin:1981, Sorkin:1982}.
Because only stable stellar configurations can persist in nature, the sequence of stellar models is terminated at that point.

The fact that cold, non-rotating NSs are bounded below a maximum mass ($M_\mathrm{TOV}$) is often referred to as the TOV limit, and the stable stellar models that exist with $dM/dp_c > 0$ and $p_c$ less than the central pressure at the TOV limit ($p_\mathrm{TOV}$) are called a \emph{stable branch} of the $M$-$R$ curve.

However, in general, the EoS can support multiple stable branches.
Famously, EoS with strong phase transitions can yield stellar sequences which lose stability when $p_c$ reaches the onset of the phase transition and regain stability when $p_c$ is slightly above the end of the phase transition~\citep{Essick:2023}.
This can lead to multiple stars with the same $M$ but different $R$ (\emph{twin stars}).
Fig.~\ref{fig:mass-radius} shows a few examples.
See, e.g.,~\citet{Essick:2023, Alford:2004, Alford:2013} and~\citet{Mroczek:2023} for more discussion.

Importantly, the naive termination condition for the TOV sequence fails to capture the physics of phase transitions (i.e., the presence of twin stars).
It is clear, then, that one must continue to solve the TOV equations for $p_c$ above the first point where $dM/dp_c=0$.
This begs the question of when, if ever, one can be confident that they have reached the end of the final stable branch.
This letter provides such a termination condition based on the stability of high-density stars: i.e., NSs with $p_c$ larger than the $p_c$ at which the $dM/dp_c = 0$ and $dR/dp_c > 0$ (n=1 radial mode loses stability) are almost certainly unstable.
This criterion automatically detects the end of all stable NS branches for EoS that support massive stars ($M_\mathrm{TOV} \gtrsim 2\,M_\odot$).\footnote{See Appendix~\ref{sec:appendix} for counter-examples with EoS that do not support massive stars.}

Additionally, I provide an efficient algorithm to construct stellar sequences.
Modern inference schemes for the EoS can require $10^4$-$10^6$ EoS proposals.
Even modest reductions in the number of stellar models required per EoS can result in significant overall savings, and my algorithm achieves \result{an order of magnitude reduction in computational cost per EoS} compared to other algorithms with, e.g., uniform grid placements in $\ln p_c$.
Accurate interpolators for macroscopic relations can be achieved in \result{$O(500)$ ms/EoS} on a modern laptop, making it possible to analyze large EoS sets with minimal computational resources in only a few hours.


\section{TOV termination conditions}
\label{sec:termination conditions}

As shown in Fig.~\ref{fig:mass-radius}, the high-density TOV sequence always approaches a fixed point
\begin{equation}
    M_\circlearrowleft \equiv \lim\limits_{p_c \rightarrow \infty} M
\end{equation}
through a counterclockwise spiral in the $M$-$R$ plane.\footnote{Similar behavior can also appear in Newtonian gravity~\citep{Bonnor:1956, Ebert:1955}.}
The location of the spiral depends on the EoS, but it always contains many local extrema in $M$.
The relevant ``static'' stability criteria~\citep{Bardeen:1966, Harrison:1965, Sorkin:1981, Sorkin:1982} can be summarized as follows: a $M$-$R$ curve that bends counterclockwise corresponds to a mode losing stability, while one that bends clockwise corresponds to a mode regaining stability.
As discussed extensively in~\citet{Harrison:1965}, the fact that $dM/dp_c > 0$ for some parts of the spiral does not guarantee stability, and instead each subsequent extremum of $M$ within the spiral corresponds to another stellar eigenmode becoming unstable in order of increasing number of radial nodes ($\mathrm{n}$).
\citet{Harrison:1965} even derives analytic expressions for the $M$-$p_c$ curve when $c_s^2$ is a constant at large $p$: $M$ is a damped-sinusoid in $\ln p_c$, and there is an infinite sequence of branches where $dM/dp_c > 0$ above $p_\mathrm{TOV}$.

Fig.~\ref{fig:mass-radius} shows the behavior of $M$-$R$, $M$-$\Lambda$ (tidal deformability), and $M$-$p_c$ for several EoS with different high-density behavior.
Annotations label which modes change stability at each local extremum in $M$.
I discuss some of the qualitative differences in these $M$-$R$ curves and their relationship to the high-density EoS below.

For EoS that support massive stars ($M_\mathrm{TOV} \gtrsim 2\,M_\odot$) and at very large $p_c$, I find that once the first harmonic ($\mathrm{n}=1$) becomes unstable, the star can never become completely stable again.
The $\mathrm{n}=1$ mode may momentarily restabilize (in particular, this occurs if a high-$p$ phase transition caused $\mathrm{n}=1$ to lose stability and $c_s^2 = 1$ above the transition; panel E of Fig.~\ref{fig:mass-radius}), but the fundamental mode ($\mathrm{n}=0$) will not restabilize even if the EoS is causal ($c_s^2 = 1$).
Fig.~\ref{fig:mass-radius} shows several such extensions.
The red curve in the top row extends the orange curve with $c_s^2=1$ at pressures above the $p_c$ at which $\mathrm{n}=1$ loses stability ($p_{n=1}$) in the orange curve.
The light-blue curve in the bottom row modifies the orange curve from the top row by introducing a $1^\mathrm{st}$-order phase transition (which causes $\mathrm{n}=1$ to lose stability) and has $c_s^2=1$ at higher densities (which causes $\mathrm{n}=1$ to regain stability, but $\mathrm{n}=0$ remains unstable).
Appendix~\ref{sec:appendix} discusses this in more detail.

In the presence of twin stars, one may see several extrema in $M$ corresponding to $\mathrm{n}=0$ losing and regaining stability.
Nevertheless, at $p_c$ higher than the $p_c$ where $\mathrm{n}=0$ becomes unstable ``for good,'' the next mode to become unstable is $\mathrm{n}=1$.
This corresponds to a local minimum in $M$ with $dR/dp_c > 0$.
At this point, additional extrema never correspond to $\mathrm{n}=0$ regaining stability.
Said another way, while $\mathrm{n}=0$ can regain stability at high $p_c$ when $\mathrm{n}=1$ is still stable, it cannot once $\mathrm{n}=1$ is unstable for massive NSs.
If one detects an unstable $\mathrm{n}=1$, then, this signals the final spiral towards $M_\circlearrowleft$ and that all stars with larger $p_c$ are unstable.

This criterion is motivated by a survey of possible high-density extensions to reasonable EoS (Fig.~\ref{fig:mass-radius}).
I do not know of a general proof, and counterexamples exist for low-$C$ stars: e.g., white dwarfs~\citep[WD; ][]{Baym:1971}, strange dwarfs~\citep{Alford:2017, Goncalves:2023, DiClemente:2024}, and EoS with $M_\mathrm{TOV} \lesssim 1.0\,M_\odot$ (Appendix~\ref{sec:appendix}).
The criterion is physically reasonable, though, as restabilization at $p_c \geq p_{n=1}$ is only possible if $c_s^2$ increases dramatically and is associated with an increase in $C$.
EoS with large $M_\mathrm{TOV}$ already have large $c_s^2$, and they are nearly as compact as BHs.
Therefore, restabilization at $p_c \geq p_{n=1}$ does not seem possible for EoSs with large $M_\mathrm{TOV}$ because both $c_s^2$ and $C$ are bounded from above by causality and BH spacetimes (formation of a trapped surface), respectively.

At low $p_c$, one also expects a minimum NS mass associated with $\mathrm{n}=0$ regaining stability after the WD branch loses stability~\citep{Harrison:1965}.
This manifests as a local minimum in $M$ with $dR/dp_c < 0$.
Typically, this minimum occurs at $p_c$ low enough to be within the NS crust (i.e., the whole star is ``crust''), where the physics is relatively well understood and one does not have to worry about exotic behavior.


\section{Adaptive TOV sequences}
\label{sec:adaptive grid}

Given stopping criteria for the TOV sequence at both high and low $p_c$, one must determine how to efficiently construct an array of $p_c$ that yields useful $M$-$R$ and $M$-$\Lambda$ curves.
The following adaptive algorithm accomplishes this by placing additional stellar models only where errors in the interpolated curves are large.

The algorithm operates via recursive bisection.
Given an initial min- and max-$p_c$, along with corresponding stellar models, it generates a new $p_c$ (mid-$p_c$, the geometric mean of min- and max-$p_c$), solves for the associated stellar model, and checks the accuracy of a linear interpolator (based on the min- and max-$p_c$) at mid-$p_c$.\footnote{More complicated interpolators could be used, but the speed and simplicity of linear interpolation work well in practice.}
If the interpolator passes a relative error tolerance, the algorithm terminates and returns the stellar models at min-, mid-, and max-$p_c$.
Otherwise, it returns the union of recursive call on the two segments defined from min- to mid-$p_c$ and from mid- to max-$p_c$.
In this way, additional models are only generated where the corresponding interpolator has relatively poor performance.
Compared to a grid with models placed uniformly in $\ln p_c$, I see as much as a \result{factor of $\sim 10$ reduction} in the number of models needed to achieve the same relative interpolator error at all points along $M$-$R$ curves.

This algorithm converges best when the function is reasonably smooth between min- and max-$p_c$, although it reliably detects kinks and bends as well.
As such, it can be seeded with an initial coarse grid in $p_c$ and iteratively applied to each segment.\footnote{The initial grid could contain only two points, though, in which case the adaptive grid would be constructed throughout the entire domain of stellar models.}

After the recursive routine terminates, the algorithm checks whether the termination conditions at high $p_c$ ($\mathrm{n}=1$ mode loses stability) and low $p_c$ ($\mathrm{n}=0$ mode regains stability) are observed within the existing set of stellar models.\footnote{Importantly, the algorithm is intended to work for NS branches and relies on the user to seed it with an appropriate initial grid of pressures.}
If not, it extends the range of $p_c$ geometrically as needed and calls the recursive routine on the new segments.
This is repeated until the termination conditions are observed (or the algorithm reaches values of $p_c$ larger than any $p$ recorded within the tabulated EoS).

Importantly, when extending the set of stellar models to lower $p_c$, the algorithm only looks for the termination condition within the new set of low-$p_c$ models.
Otherwise, it may prematurely exit if there are multiple stable branches at higher $p_c$.
Fig.~\ref{fig:mass-radius} shows such a EoS with a phase transition at low densities (just above $n_\mathrm{sat}$) that yields a short unstable branch at $M \sim 0.25\,M_\odot$.
Again, to reliably identify the presence of such features, one must instantiate the search for the end of low-$p_c$ stable branches at $n_c \lesssim n_\mathrm{sat}$, often well within the crust.

This issue does not arise when extending the sequence to higher $p_c$ as long as one focuses on the range of $p_c$ relevant for NSs ($p_c \gtrsim \mathrm{MeV}/\mathrm{fm}^3$) and stays well above what is relevant for WDs ($p_c \sim 6\times10^{-7}\,\mathrm{MeV}/\mathrm{fm}^3$). 

In principle, this algorithm could be instantiated with a single $p_c$ and rely on the extension prescriptions to both higher and lower $p_c$.
However, the geometric expansion \textit{de facto} results in less-well-optimized grid placement than allowing the adaptive algorithm to choose where to add more grid points throughout a wider initial range.
I find that beginning with a reasonable initial guess for the min- and max-$p_c$ requires \result{$O(10\%)$ fewer} stellar models compared to relying only on the geometric expansion.

Altogether, my implementation~\citep{universality} requires \result{$O(3)$ ms/model} on a \result{modern laptop ($3\,$GHz 11th Gen Intel i7-1185G7)}.
Reliable interpolators for $M$-$R$ can be constructed with with \result{$\sim$100-150 models (interpolators for $M$-$\Lambda$ require more points; see discussion below)}, yielding an runtime of \result{$\lesssim O(500)$ ms/EoS}.\footnote{It takes longer to read the EoS table from and write the TOV solutions to disk.}

However, a significant fraction of the stellar models produced have low $p_c$ (large $R$ and $\Lambda$).
If one is only interested in, e.g., stars with $M \gtrsim 0.5\,M_\odot$, then many of the models at lower $p_c$ can be skipped.
The reduced range may need \result{$\lesssim 50$ models} for accurate interpolation, but this requires user expertise to guarantee that min-$p_c$ is chosen appropriately.

Finally, solving for only ($M$, $R$) takes \result{$O(3)$ ms/model} whereas solving for ($M$, $R$, $\Lambda$) takes \result{$O(6)$ ms/model}.
Additionally, the adaptive grid often requires fewer $p_c$ to converge for just $M$-$R$ compared to $M$-$\Lambda$.
The difference can be as large as \result{a factor of $\sim$5}, but more commonly is a factor of \result{$\lesssim 2$}.
As such, it is not uncommon to find \result{a combined factor of several speedup} when only solving for ($M$, $R$). 
One might achieve a net speedup by first solving for ($M$, $R$) with a coarse interpolator to identify the relevant range of $p_c$ and then re-solving with a finer resolution for ($M$, $R$, $\Lambda$) only within the relevant range.
This may be particularly helpful if some EoS will be discarded based on information available from the $M$-$R$ curve, like $M_\mathrm{TOV}$.
In this case, one would never have to solve for $M$-$\Lambda$ for some EoS.
However, I leave a precise quantification of the potential speedup to future work.

Equipped with this algorithm for rapid and reliable TOV sequences, I now explore exotic behavior at both low- and high-$p_c$.


\begin{figure*}
    \includegraphics[width=1.0\textwidth]{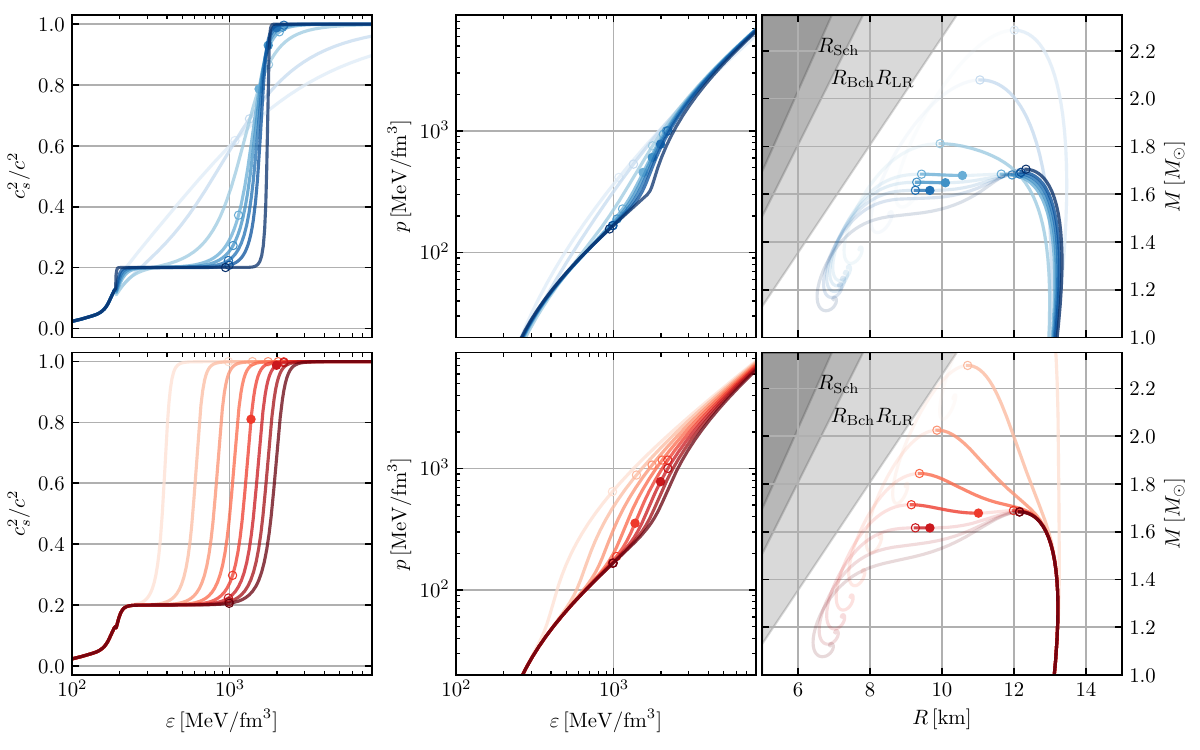}
    \caption{
        EoS with monotonically increasing $c_s^2$ that produce twin stars: (\emph{left}) $c_s^2$ vs. $\varepsilon${, (\emph{center}) $p$ vs. $\varepsilon$,} and (\emph{right}) $M$ vs. $R$.
        Stable and unstable branches are denoted as in Fig.~\ref{fig:mass-radius}.
        The ability of such EoS to support twin stars is a combination of (\emph{top}) the smoothness of $c_s^2$ and (\emph{bottom}) the size of the density range where $c_s^2$ is nearly constant.
        This behavior is reminiscent of $2^\mathrm{nd}$-order phase transitions as there is a (near) discontinuity in $dc_s^2/d\varepsilon = d^2p/d\varepsilon^2$ rather than $c_s^2 = dp/d\varepsilon$.
        Although not shown here, EoS that support twin stars do so even if $c_s^2 \rightarrow 1/3$ at higher densities.
        I use examples with $M_\mathrm{TOV} < 2\,M_\odot$ to more clearly show the variety of behavior.
        Fig.~\ref{fig:mass-radius} (panel B) shows an example with $M_\mathrm{TOV} \sim 2\,M_\odot$.
    }
    \label{fig:twin example}
\end{figure*}

\section{Exotic low-density behavior}
\label{sec:low-density exotica}

As briefly discussed before, the EoS can support multiple stable branches at low $p_c$ while still matching \textit{ab initio} theory \citep{Keller:2023} at $n \lesssim n_\mathrm{sat}$, and similar behavior has been reported elsewhere~\citep[see, e.g., Category IV EoSs in][]{Christian:2017}.
In particular, there can be a phase transition just above $n_\mathrm{sat}$ that is consistent with astrophysical observations, which currently allow phase transitions (and multiple stable branches) at $M \lesssim M_\odot$ and/or $M \gtrsim 2\,M_\odot$~\citep{Legred:2021, Essick:2023}.
In fact, although far from being certain, if a large slope parameter at $n_\mathrm{sat}$ ($L \gtrsim 100\,\mathrm{MeV}$) is observed experimentally, such a phase transition just above $n_\mathrm{sat}$ could actually be preferred by current observations~\citep{Essick:2021a, Essick:2021b}.

Fig.~\ref{fig:mass-radius} shows an example of such an EoS.
Often, small $R(1.4\,M_\odot)$ are associated with this behavior; only the EoS with a low-$p$ phase transition supports $R(1.4\,M_\odot) < 10\,\mathrm{km}$ in this (admittedly small) sample.


\section{Exotic high-density behavior}
\label{sec:high-density exotica}

\begin{figure*}
    \includegraphics[width=1.0\textwidth]{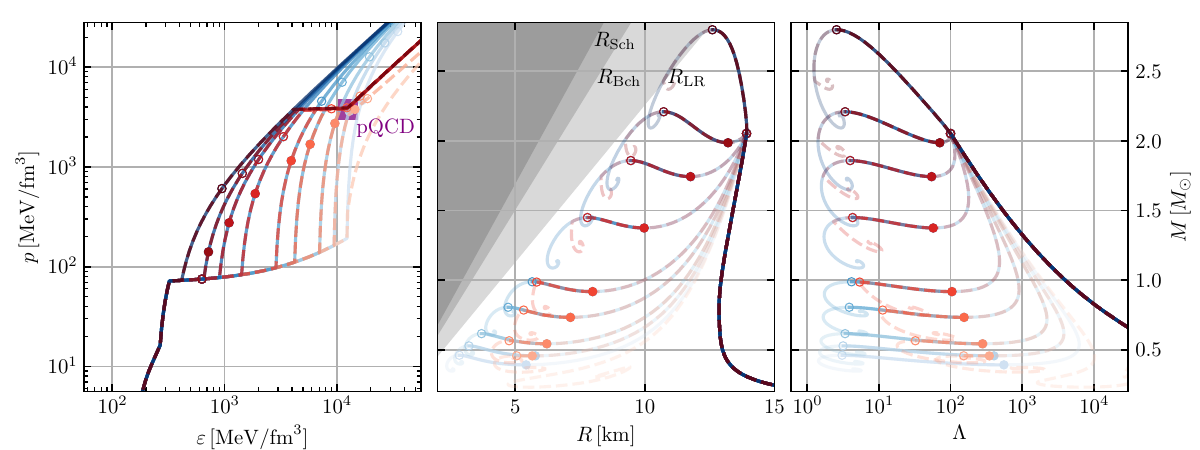}
    \caption{
        EoS with high-density phase transitions and HiPO branches: (\emph{left}) $p$ vs. $\varepsilon$, (\emph{center}) $M$ vs. $R$, and (\emph{right}) $\Lambda$ vs $M$.
        Stability is indicated as in Figs.~\ref{fig:mass-radius} and~\ref{fig:twin example}.
        I show (\emph{solid lines}) EoS with $c_s^2 = 1$ up to arbitrarily high densities and (\emph{dashed lines}) EoS that match (\emph{purple square}) approximate pQCD predictions at \externalresult{$\varepsilon \approx 12.5\,\mathrm{GeV}/\mathrm{fm}^3$} for $p$ and $c_s^2$~\citep{Komoltsev:2023}.
        The single EoS that fails to support a HiPO branch does so because $c_s^2 = 1/3$ at all densities above its phase transition (i.e., $c_s^2$ never reaches $\sim 1$).
        This is also the only EoS which fails to match the pQCD prediction, suggesting that pQCD places an effective lower limit on HiPO masses of $\gtrsim 0.5\,M_\odot$.
    }
    \label{fig:big trans examples}
\end{figure*}

Fig.~\ref{fig:mass-radius} also shows various high-density behavior, including stiff ($c_s^2 \sim 1$) and soft ($c_s^2 \sim 1/3$) EoS both with and without phase transitions.
All stellar sequences eventually end in a counterclockwise spiral.
However, the path by which the high-density models enter this spiral can vary.
I focus on EoS that approximately satisfy current observations constraints (i.e., \result{follow $\chi$EFT at low densities and have $R(1.4\,M_\odot) \sim 12.5\,\mathrm{km}$, $M_\mathrm{TOV} \gtrsim 2\,M_\odot$}).
See Appendix~\ref{sec:appendix} for discussion of more general behavior.

The vanilla behavior for EoS with $M_\mathrm{TOV} \gtrsim 2\,M_\odot$ is as follows: $\mathrm{n}=0$ becomes unstable, then $M$ decreases and enters a spiral.
The spiral is always centered on $M_\circlearrowleft < M_\mathrm{TOV}$, and the $M$-$R$ curves never cross themselves (i.e., \emph{identical} twin stars are not physical).
It is interesting that EoS with $c_s^2 \sim 1$ for $p \sim p_\mathrm{TOV}$ can produce $R_\mathrm{TOV}$ comparable to (and smaller than!) their light-ring ($R_\mathrm{LR} = 3M$).
This could create a (small) resonating cavity like those invoked to motivate searches for gravitational-wave echoes~\citep{Micchi:2021, Oshita:2019}.

The presence of twin stars only moderately complicates this behavior: $\mathrm{n}=0$ can regain stability, and the second stable branch can, but does not have to, reach higher $M$ than the first stable branch.
Again, later stable branches with $c_s^2 \sim 1$ can terminate near $R_\mathrm{LR}$.
See~\citet{Alford:2013} and~\citet{Christian:2017} for categorization schemes for twin-star branches.

If a phase transition occurs at $p > p_\mathrm{TOV}$, it can still cause a change in stability.
However, $\mathrm{n}=1$ loses and regains stability rather than $\mathrm{n}=0$.
In this case, the star as a whole remains unstable (panel E in Fig.~\ref{fig:mass-radius}).

Counter to claims made in the literature, it is also possible to generate twin stars \emph{without} a discontinuity or decrease in $c_s^2$ (Fig.~\ref{fig:twin example}).
It is well-known that EoS with $1^\mathrm{st}$-order phase transitions, either with mixed phases~\citep{Han:2019, Glendenning:1998, Schertler:2000, Alvarez-Castillo:2014} or without~\citep[e.g.,][]{Alford:2013}, can support twin stars.
However, all these EoS contain non-monotonic $c_s^2$, and $c_s^2$ decreases (often sharply) at the onset of the phase transition.
Instead, I find that even EoS with monotonically increasing $c_s^2$ can produce twin stars.\footnote{For these Eos, $c_s^2$ behaves as one might expect for a 2$^\mathrm{nd}$-order phase transition.}
This phenomenology is difficult (but not impossible!) to achieve while satisfying current observational constraints (panel B in Fig.~\ref{fig:mass-radius}).
Relaxing the requirement that $M_\mathrm{TOV} \gtrsim 2\,M_\odot$ allows a much wider range of EoS with monotonic $c_s^2$ to produce twin stars.
In Fig.~\ref{fig:twin example}, all EoS have $c_s^2$ that monotonically increase with $\varepsilon$ and correspondingly smooth $p(\varepsilon)$.
Nevertheless, rapid increases in $c_s^2$ at high densities can support twin stars even though $c_s^2$ does not decrease at the end of the first stable branch.

Finally, phase transitions at high $p$ with large latent energies can generate stable branches with $M \ll M_\mathrm{TOV}$ and $R \ll R_\mathrm{TOV}$ if $c_s^2 \sim 1$ beyond the transition (Figs.~\ref{fig:mass-radius} and~\ref{fig:big trans examples}).
These stars can have central pressures much higher than $p_\mathrm{TOV}$, often $\gtrsim \mathrm{GeV}/\mathrm{fm}^3$.
I will refer to them as high-$p_c$ objects, or HiPOs.
Without sufficient exploration of high-$p_c$ TOV solutions, these branches could have easily been missed (i.e., there is a long unstable branch before the sequence regains stability).

HiPO branches appear to be common if $c_s^2 \sim 1$ is allowed at large $p$.
In fact, it may be possible to extend most EoS that satisfy astrophysical constraints with this type of high-density behavior, and recent proposals for how to incorporate theoretical predictions from perturbative quantum chromodynamics~\citep[pQCD; ][]{Komoltsev:2021, Komoltsev:2023} invoke EoS like these.
One can also tune the high-$p$ EoS to place $M_\circlearrowleft$ at nearly any value $\lesssim M_\mathrm{TOV}$.
Reducing either the latent energy or density at which $c_s^2 \rightarrow 1/3$ tends to increase $M_\circlearrowleft$ and decrease the extent of the stable branch.
Fig.~\ref{fig:big trans examples} shows the typical behavior.

It has not been previously shown that such stars can reach sub-solar masses, although~\citet{Christian:2017} would classify HiPOs as Category II EoS, and~\citet{Alford:2013} would call them Type D (``disconnnected'').
Furthermore, one may be concerned that such EoS require large enough latent energies that their pressures are incompatible with predictions from pQCD~\citep[$p \sim 4\,\mathrm{GeV}/\mathrm{fm}^3$ at $\varepsilon \sim 12.5\,\mathrm{GeV}/\mathrm{fm}^3$; ][]{Komoltsev:2023}.
However, Fig.~\ref{fig:big trans examples} and Sec.~\ref{sec:conclusions} show that such EoSs can be compatible with pQCD, although matching pQCD does limit the lowest HiPO masses to $\gtrsim 0.5\,M_\odot$.


\section{Summary}
\label{sec:conclusions}

Based on the stability of dense stars, I introduce improved termination conditions for sequences of stellar models.
Together with an efficient algorithm to construct reliable interpolators, I explored the types of behaviors observed at low- and high-$p_c$.
Surprisingly, I find that
\begin{itemize}
    \item an EoS can support twin stars even though $c_s^2$ always increases monotonically, and
    \item large phase transitions at high $p$ can produce stable (HiPO) branches at much smaller ($M$, $R$) than the TOV limit with $p_c$ near what is directly accessible with pQCD.
\end{itemize}
Both behaviors are possible while maintaining $M_\mathrm{TOV} \gtrsim 2\,M_\odot$ and reasonable radii.
Their existence shows that more care may be needed in the literature\footnote{Or, at least, within my body of work~\citep{Essick:2019, Essick:2020, Essick:2021a, Essick:2021b, Essick:2023}.} when making claims that, e.g., the presence of twin stars robustly implies the presence of a phase transition (which manifests as a non-monotonic $c_s^2$) or that the presence of a sub-solar mass object with small $\Lambda$ (observationally indistinguishable from a BH with $\Lambda_\mathrm{BH} = 0$) is clear evidence for sub-solar mass BHs~\citep{Crescimbeni:2024, Golomb:2024}.
Instead, twin stars near $M_\mathrm{TOV}$ do not require $c_s^2$ to decrease, and there can be stable NS branches with $\Lambda \sim O(10)$ and $M \lesssim M_\odot$.\footnote{As such, more care is also warranted with statements that $p > p_\mathrm{TOV}$ is both ``unobserved and unobservable'' \citep{Essick:2023}, as the presence and/or absence of such branches directly constraints the high-$p$ EoS.}

That being said, even if HiPO branches exist, the range of masses a single EoS can support is limited.
While it may be possible to produce an EoS that matches the mass of any individual object with $\Lambda \sim 0$, it may be difficult to explain the observation of a population of compact objects over a range of $M$.
Alternative explanations would not be ruled out, though, including complex dark sectors that support ``mirror stars''~\citep{Hippert:2022}.

Taking the HiPO branches at face value, let us consider how one might form such a star dynamically.
The jump from the ``normal'' NS branch ($p < p_\mathrm{TOV}$) to the HiPO branch is in many ways analogous to the jump from the WD branch to the NS branches.
There is a reduction in the number of particles ($N$) within the star\footnote{See the discussion of ``pumping up the central density'' in \citet{Harrison:1965}.}
\begin{equation}
    N = \int\limits_0^R dr \frac{4\pi r^2 n}{\sqrt{1 - 2m/r}}
\end{equation}
implying that a star collapsing from $M_\mathrm{TOV}$ to the HiPO branch would have to shed mass, as much as $O(1)\,M_\odot$.
This could happen in a process analogous to a core-collapse supernova (CCSN),\footnote{See also quark nova~\citep{Ouyed:2002, Jaikumar:2007}, which similarly posit a transition from a NS to a more compact strange star (but at much lower densities), and discussion of ``QCD-phase transitions'' as an explosion mechanism for CCSN~\citep{Heinimann:2016}.} although it is unclear whether the HiPO branch could actually launch a shock in the same way that a proto-NS is thought to within CCSN.
The infalling matter could overwhelm the HiPO branch and cause direct collapse to a BH.
Also, Fig.~\ref{fig:big trans examples} shows that some HiPO branches have $R \sim 2M_\mathrm{TOV}$, meaning that the star would have to shed a significant amount of energy within the dynamical timescale of the collapse to avoid forming a trapped surface.
Detailed simulations will likely be needed to resolve these questions.

Interestingly, if one assumes the rest-mass per particle is the same on both the normal and HiPO branches, essentially all of the HiPO branches are less bound per particle than the normal branch ($(Nm_{n} - M)/N$ is smaller).
However, if the particles produced in the phase transition instead have a rest-mass comparable to the chemical potential at the onset of the transition~\citep[$\mu_\mathrm{onset} \sim 1.3 m_{n}$ in Fig.~\ref{fig:big trans examples}, not dissimilar to some excited baryonic states;][]{Workman:2022} and that most of the particles in HiPOs are of the exotic type, then the HiPO branch can be much more tightly bound per particle than any normal NS.
This is, again, analogous to the way that WDs collapse to NSs: the electron is lighter than a nucleon, and NSs are more tightly bound than WDs.

Additionally, while Fig.~\ref{fig:big trans examples} shows that HiPOs with masses as low as $0.5\,M_\odot$ can be compatible with pQCD, the latent energy achievable may also be limited by the degeneracy pressure of the new particle.
That is, high densities of degenerate particles still contribute some degeneracy pressure (a lower limit on the partial pressure), and this imposes an upper limit on the allowed latent energy before the total pressure beings to increase significantly.
I leave a precise quantification to future work.

Finally, taking the collapse scenario seriously, I estimate the total amount of energy that would be released as the difference between the initial and final states.
Some (very) rough bounds can be set by assuming that the rest-mass from the difference in particle numbers between the TOV star and the HiPO branch is radiated as nucleons or that no rest-mass is radiated:
\begin{equation}
    M_\mathrm{TOV} - (M + (N_\mathrm{TOV}-N)m_{n}) \leq \Delta E \leq M_\mathrm{TOV} - M
\end{equation}
Depending on the high-$p$ EoS, this yields $\Delta E \sim 0.2$-$0.6\,M_\odot \sim 10^{53}$-$10^{54}\,\mathrm{ergs}$, which is larger than typical CCSN, but only by a few orders of magnitude.

It may be plausible, then, that a subpopulation of electromagnetic transients could in fact be the collapse from normal NSs to HiPOs.
The ejected mass would be considerably lower than for CCSN, though, as there is much less initial stellar material.
However, it may have a very large kinetic energy.
Additionally, the ejected mass would likely be very neutron rich, which may yield light-curves closer to a kilonova than a CCSN (faster decay rate, dimmer, and redder as radiation may be trapped by the high-opacity from $r$-process elements).
The rate of such events could be quite low, though, as not all NSs are likely to undergo this process.


\begin{acknowledgments}

I am deeply indebted to Chris Matzner, Bob Wald and Daniel Holz for discussions and suggested reading during the preparation of this manuscript.
I also thank David Curtin and Janosz Dewberry for helpful discussions.

R.E. is supported by the Natural Sciences \& Engineering Research Council of Canada (NSERC) through a Discovery Grant (RGPIN-2023-03346).

Implementations of these algorithms are available in \texttt{universality}~\citep{universality}.
This study would not have been possible without the following software: \texttt{numpy}~\citep{numpy}, \texttt{scipy}~\citep{scipy}, and \texttt{matplotlib}~\citep{matplotlib}.

\end{acknowledgments}


\bibliography{biblio}

\begin{thebibliography}{}
\expandafter\ifx\csname natexlab\endcsname\relax\def\natexlab#1{#1}\fi
\providecommand{\url}[1]{\href{#1}{#1}}
\providecommand{\dodoi}[1]{doi:~\href{http://doi.org/#1}{\nolinkurl{#1}}}
\providecommand{\doeprint}[1]{\href{http://ascl.net/#1}{\nolinkurl{http://ascl.net/#1}}}
\providecommand{\doarXiv}[1]{\href{https://arxiv.org/abs/#1}{\nolinkurl{https://arxiv.org/abs/#1}}}

\bibitem[{Alford {et~al.}(2005)Alford, Braby, Paris, \& Reddy}]{Alford:2004}
Alford, M., Braby, M., Paris, M., \& Reddy, S. 2005, Astrophys. J., 629, 969,
  \dodoi{10.1086/430902}

\bibitem[{Alford {et~al.}(2013)Alford, Han, \& Prakash}]{Alford:2013}
Alford, M.~G., Han, S., \& Prakash, M. 2013, Phys. Rev. D, 88, 083013,
  \dodoi{10.1103/PhysRevD.88.083013}

\bibitem[{Alford {et~al.}(2017)Alford, Harris, \& Sachdeva}]{Alford:2017}
Alford, M.~G., Harris, S.~P., \& Sachdeva, P.~S. 2017, Astrophys. J., 847, 109,
  \dodoi{10.3847/1538-4357/aa8509}

\bibitem[{Alvarez-Castillo \& Blaschke(2015)}]{Alvarez-Castillo:2014}
Alvarez-Castillo, D.~E., \& Blaschke, D. 2015, Phys. Part. Nucl., 46, 846,
  \dodoi{10.1134/S1063779615050032}

\bibitem[{{Bardeen} {et~al.}(1966){Bardeen}, {Thorne}, \&
  {Meltzer}}]{Bardeen:1966}
{Bardeen}, J.~M., {Thorne}, K.~S., \& {Meltzer}, D.~W. 1966, \apj, 145, 505,
  \dodoi{10.1086/148791}

\bibitem[{{Baym} {et~al.}(1971){Baym}, {Pethick}, \& {Sutherland}}]{Baym:1971}
{Baym}, G., {Pethick}, C., \& {Sutherland}, P. 1971, \apj, 170, 299,
  \dodoi{10.1086/151216}

\bibitem[{Bonnar(1956)}]{Bonnor:1956}
Bonnar, W.~B. 1956, Monthly Notices of the Royal Astronomical Society, 116,
  351, \dodoi{10.1093/mnras/116.3.351}

\bibitem[{Christian {et~al.}(2018)Christian, Zacchi, \&
  Schaffner-Bielich}]{Christian:2017}
Christian, J.-E., Zacchi, A., \& Schaffner-Bielich, J. 2018, Eur. Phys. J. A,
  54, 28, \dodoi{10.1140/epja/i2018-12472-y}

\bibitem[{Crescimbeni {et~al.}(2024)Crescimbeni, Franciolini, Pani, \&
  Riotto}]{Crescimbeni:2024}
Crescimbeni, F., Franciolini, G., Pani, P., \& Riotto, A. 2024.
\newblock \doarXiv{2402.18656}

\bibitem[{Cromartie {et~al.}(2019)}]{Cromartie:2019}
Cromartie, H.~T., {et~al.} 2019, Nature Astron., 4, 72,
  \dodoi{10.1038/s41550-019-0880-2}

\bibitem[{Di~Clemente {et~al.}(2024)Di~Clemente, Drago, \&
  Pagliara}]{DiClemente:2024}
Di~Clemente, F., Drago, A., \& Pagliara, G. 2024.
\newblock \doarXiv{2406.03137}

\bibitem[{{Ebert}(1955)}]{Ebert:1955}
{Ebert}, R. 1955, Zeitschrift für Astrophysik, 37, 217

\bibitem[{Essick(2024)}]{universality}
Essick, R. 2024, Universality (2024-08-06),  Zenodo,
  \dodoi{10.5281/zenodo.13241272}

\bibitem[{Essick {et~al.}(2020{\natexlab{a}})Essick, Landry, \&
  Holz}]{Essick:2019}
Essick, R., Landry, P., \& Holz, D.~E. 2020{\natexlab{a}}, Phys. Rev. D, 101,
  063007, \dodoi{10.1103/PhysRevD.101.063007}

\bibitem[{Essick {et~al.}(2021{\natexlab{a}})Essick, Landry, Schwenk, \&
  Tews}]{Essick:2021b}
Essick, R., Landry, P., Schwenk, A., \& Tews, I. 2021{\natexlab{a}}, Phys. Rev.
  C, 104, 065804, \dodoi{10.1103/PhysRevC.104.065804}

\bibitem[{Essick {et~al.}(2023)Essick, Legred, Chatziioannou, Han, \&
  Landry}]{Essick:2023}
Essick, R., Legred, I., Chatziioannou, K., Han, S., \& Landry, P. 2023, Phys.
  Rev. D, 108, 043013, \dodoi{10.1103/PhysRevD.108.043013}

\bibitem[{Essick {et~al.}(2020{\natexlab{b}})Essick, Tews, Landry, Reddy, \&
  Holz}]{Essick:2020}
Essick, R., Tews, I., Landry, P., Reddy, S., \& Holz, D.~E. 2020{\natexlab{b}},
  Phys. Rev. C, 102, 055803, \dodoi{10.1103/PhysRevC.102.055803}

\bibitem[{Essick {et~al.}(2021{\natexlab{b}})Essick, Tews, Landry, \&
  Schwenk}]{Essick:2021a}
Essick, R., Tews, I., Landry, P., \& Schwenk, A. 2021{\natexlab{b}}, Phys. Rev.
  Lett., 127, 192701, \dodoi{10.1103/PhysRevLett.127.192701}

\bibitem[{Fonseca {et~al.}(2021)}]{Fonseca:2021}
Fonseca, E., {et~al.} 2021, Astrophys. J. Lett., 915, L12,
  \dodoi{10.3847/2041-8213/ac03b8}

\bibitem[{Glendenning \& Kettner(2000)}]{Glendenning:1998}
Glendenning, N.~K., \& Kettner, C. 2000, Astron. Astrophys., 353, L9.
\newblock \doarXiv{astro-ph/9807155}

\bibitem[{Golomb {et~al.}(2024)Golomb, Legred, Chatziioannou, Abac, \&
  Dietrich}]{Golomb:2024}
Golomb, J., Legred, I., Chatziioannou, K., Abac, A., \& Dietrich, T. 2024.
\newblock \doarXiv{2403.07697}

\bibitem[{Goncalves {et~al.}(2023)Goncalves, Jimenez, \&
  Lazzari}]{Goncalves:2023}
Goncalves, V.~P., Jimenez, J.~C., \& Lazzari, L. 2023, Eur. Phys. J. A, 59,
  251, \dodoi{10.1140/epja/s10050-023-01175-5}

\bibitem[{Han {et~al.}(2019)Han, Mamun, Lalit, Constantinou, \&
  Prakash}]{Han:2019}
Han, S., Mamun, M. A.~A., Lalit, S., Constantinou, C., \& Prakash, M. 2019,
  Phys. Rev. D, 100, 103022, \dodoi{10.1103/PhysRevD.100.103022}

\bibitem[{Harris {et~al.}(2020)Harris, Millman, van~der Walt, Gommers,
  Virtanen, Cournapeau, Wieser, Taylor, Berg, Smith, Kern, Picus, Hoyer, van
  Kerkwijk, Brett, Haldane, del R{\'{i}}o, Wiebe, Peterson,
  G{\'{e}}rard-Marchant, Sheppard, Reddy, Weckesser, Abbasi, Gohlke, \&
  Oliphant}]{numpy}
Harris, C.~R., Millman, K.~J., van~der Walt, S.~J., {et~al.} 2020, Nature, 585,
  357, \dodoi{10.1038/s41586-020-2649-2}

\bibitem[{{Harrison} {et~al.}(1965){Harrison}, {Thorne}, {Wakano}, \&
  {Wheeler}}]{Harrison:1965}
{Harrison}, B.~K., {Thorne}, K.~S., {Wakano}, M., \& {Wheeler}, J.~A. 1965,
  {Gravitation Theory and Gravitational Collapse}

\bibitem[{Heinimann {et~al.}(2016)Heinimann, Hempel, \&
  Thielemann}]{Heinimann:2016}
Heinimann, O., Hempel, M., \& Thielemann, F.-K. 2016, Phys. Rev. D, 94, 103008,
  \dodoi{10.1103/PhysRevD.94.103008}

\bibitem[{Hippert {et~al.}(2022)Hippert, Setford, Tan, Curtin, Noronha-Hostler,
  \& Yunes}]{Hippert:2022}
Hippert, M., Setford, J., Tan, H., {et~al.} 2022, Phys. Rev. D, 106, 035025,
  \dodoi{10.1103/PhysRevD.106.035025}

\bibitem[{Hunter(2007)}]{matplotlib}
Hunter, J.~D. 2007, Computing in Science \& Engineering, 9, 90,
  \dodoi{10.1109/MCSE.2007.55}

\bibitem[{{Jaikumar} {et~al.}(2007){Jaikumar}, {Meyer}, {Otsuki}, \&
  {Ouyed}}]{Jaikumar:2007}
{Jaikumar}, P., {Meyer}, B.~S., {Otsuki}, K., \& {Ouyed}, R. 2007, Astronomy
  and Astrophysics, 471, 227, \dodoi{10.1051/0004-6361:20066593}

\bibitem[{Kastaun \& Ohme(2024)}]{Kastaun:2024}
Kastaun, W., \& Ohme, F. 2024.
\newblock \doarXiv{2404.11346}

\bibitem[{Keller {et~al.}(2023)Keller, Hebeler, \& Schwenk}]{Keller:2023}
Keller, J., Hebeler, K., \& Schwenk, A. 2023, Phys. Rev. Lett., 130, 072701,
  \dodoi{10.1103/PhysRevLett.130.072701}

\bibitem[{Komoltsev \& Kurkela(2022)}]{Komoltsev:2021}
Komoltsev, O., \& Kurkela, A. 2022, Phys. Rev. Lett., 128, 202701,
  \dodoi{10.1103/PhysRevLett.128.202701}

\bibitem[{Komoltsev {et~al.}(2023)Komoltsev, Somasundaram, Gorda, Kurkela,
  Margueron, \& Tews}]{Komoltsev:2023}
Komoltsev, O., Somasundaram, R., Gorda, T., {et~al.} 2023.
\newblock \doarXiv{2312.14127}

\bibitem[{Legred {et~al.}(2021)Legred, Chatziioannou, Essick, Han, \&
  Landry}]{Legred:2021}
Legred, I., Chatziioannou, K., Essick, R., Han, S., \& Landry, P. 2021, Phys.
  Rev. D, 104, 063003, \dodoi{10.1103/PhysRevD.104.063003}

\bibitem[{Lindblom(2014)}]{Lindblom:2014}
Lindblom, L. 2014, AIP Conference Proceedings, 1577, 153,
  \dodoi{10.1063/1.4861951}

\bibitem[{Longo~Micchi {et~al.}(2021)Longo~Micchi, Afshordi, \&
  Chirenti}]{Micchi:2021}
Longo~Micchi, L.~F., Afshordi, N., \& Chirenti, C. 2021, Phys. Rev. D, 103,
  044028, \dodoi{10.1103/PhysRevD.103.044028}

\bibitem[{Mroczek {et~al.}(2023)Mroczek, Miller, Noronha-Hostler, \&
  Yunes}]{Mroczek:2023}
Mroczek, D., Miller, M.~C., Noronha-Hostler, J., \& Yunes, N. 2023.
\newblock \doarXiv{2309.02345}

\bibitem[{Oppenheimer \& Volkoff(1939)}]{Oppenheimer:1939}
Oppenheimer, J.~R., \& Volkoff, G.~M. 1939, Phys. Rev., 55, 374,
  \dodoi{10.1103/PhysRev.55.374}

\bibitem[{Oshita \& Afshordi(2019)}]{Oshita:2019}
Oshita, N., \& Afshordi, N. 2019, Phys. Rev. D, 99, 044002,
  \dodoi{10.1103/PhysRevD.99.044002}

\bibitem[{{Ouyed} {et~al.}(2002){Ouyed}, {Dey}, \& {Dey}}]{Ouyed:2002}
{Ouyed}, R., {Dey}, J., \& {Dey}, M. 2002, Astronomy and Astrophysics, 390,
  L39, \dodoi{10.1051/0004-6361:20020982}

\bibitem[{Schertler {et~al.}(2000)Schertler, Greiner, Schaffner-Bielich, \&
  Thoma}]{Schertler:2000}
Schertler, K., Greiner, C., Schaffner-Bielich, J., \& Thoma, M.~H. 2000, Nucl.
  Phys. A, 677, 463, \dodoi{10.1016/S0375-9474(00)00305-5}

\bibitem[{{Sorkin}(1981)}]{Sorkin:1981}
{Sorkin}, R. 1981, \apj, 249, 254, \dodoi{10.1086/159282}

\bibitem[{{Sorkin}(1982)}]{Sorkin:1982}
{Sorkin}, R.~D. 1982, \apj, 257, 847, \dodoi{10.1086/160034}

\bibitem[{Tolman(1939)}]{Tolman:1939}
Tolman, R.~C. 1939, Phys. Rev., 55, 364, \dodoi{10.1103/PhysRev.55.364}

\bibitem[{Virtanen {et~al.}(2020)Virtanen, Gommers, Oliphant, Haberland, Reddy,
  Cournapeau, Burovski, Peterson, Weckesser, Bright, {van der Walt}, Brett,
  Wilson, Millman, Mayorov, Nelson, Jones, Kern, Larson, Carey, Polat, Feng,
  Moore, {VanderPlas}, Laxalde, Perktold, Cimrman, Henriksen, Quintero, Harris,
  Archibald, Ribeiro, Pedregosa, {van Mulbregt}, \& {SciPy 1.0
  Contributors}}]{scipy}
Virtanen, P., Gommers, R., Oliphant, T.~E., {et~al.} 2020, Nature Methods, 17,
  261, \dodoi{10.1038/s41592-019-0686-2}

\bibitem[{Workman {et~al.}(2022)}]{Workman:2022}
Workman, R.~L., {et~al.} 2022, PTEP, 2022, 083C01, \dodoi{10.1093/ptep/ptac097}

\end{thebibliography}


\newpage
\appendix


\section{Low-Compactness Stars}
\label{sec:appendix}

\begin{figure*}
    \includegraphics[width=1.0\textwidth]{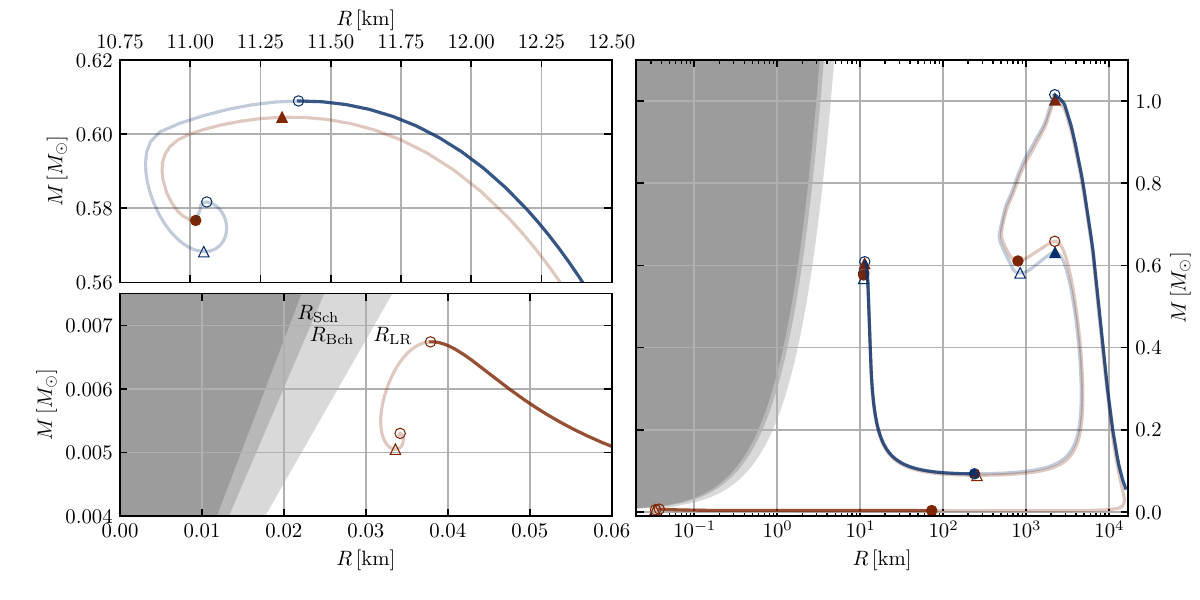}
    \caption{
        $M$-$R$ curve for an example EoS with low $M_\mathrm{TOV}$ that restabilizes after (\emph{right}) $\mathrm{n}=1$ loses stability on the WD branch and (\emph{top left}) after $\mathrm{n}=2$ loses stability on the NS branch.
        (\emph{bottom left}) It also supports a very-low-mass HiPO branch.
        Stable and unstable branches are denoted as in the other figures, as are markers for which mode loses or regains stability.
        Because the curves are similar, I also color stellar models with $p_c$ less than where $\mathrm{n}=2$ loses stability blue and models with larger $p_c$ red.
        This EoS is inconsistent with astrophysical observations, and includes a phase transition with an implausibly large latent energy, but it nonetheless shows that such behavior is not forbidden by the TOV equations.
    }
    \label{fig:unwind}
\end{figure*}

The examples in the main text focus on the behavior of compact NSs.
In general, though, the stellar sequence can regain stability after both $\mathrm{n}=0$ and $\mathrm{n}=1$ lose stability at pressures above the WD branch.
This happens at $R\sim10^3\,\mathrm{km}$ for the BPS crust~\citep{Baym:1971} used to construct all the examples in this work (Fig.~\ref{fig:unwind}), and there are many other examples in the literature, including ``strange dwarfs'' in which a phase transition occurs at pressures between the WD and NS branches~\citep{Alford:2017, Goncalves:2023, DiClemente:2024}.
Additionally, one can even construct (highly artificial) EoS that unwind the spiral and regain stability at $p_c$ above where $\mathrm{n}=2$ becomes unstable.
Fig.~\ref{fig:unwind} shows such an example, demonstrating that this behavior is not forbidden by the TOV equations for low-$C$ stellar models.

\begin{figure}
    \begin{center}
        \includegraphics[width=0.5\columnwidth]{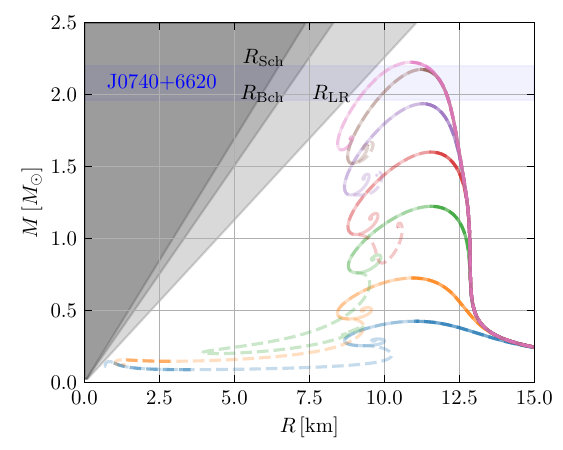}
    \end{center}
    \caption{
        $M$ - $R$ relations for a sequence of EoS with increasing $M_\mathrm{TOV}$.
        (\emph{solid}) EoS with $c_s^2$ that plateau at a constant and (\emph{dashed of the same color}) causal extensions ($c_s^2=1$) at $p \geq p_{n=1}$.
        As in other figures, I show stable branches in dark colors and unstable branches in light colors.
        For clarity, I do not label the points at which individual modes change stability, although $p_{n=1}$ is evident from the point where the solid and dashed curves diverge.
        I also show $R_\mathrm{Sch}$, $R_\mathrm{Bch}$, $R_\mathrm{LR}$, and the approximate uncertainty in the mass of J0740+6620.
        For EoS that produce $M_\mathrm{TOV} \ll M_\odot$, it is possible to restabilize the stellar sequence at $p_c \geq p_{n=1}$, but this is not possible for EoS that yield more compact stars and larger $M_\mathrm{TOV}$.
    }
    \label{fig:compactness sequence}
\end{figure}

Clearly, then, the termination condition does not work universally.
Dramatic increases in $c_s$ at low pressures can restabilize the stellar sequence, although they are associated with collapse to a much larger compactness.
However, as discussed in the main text, for high-$C$ stars ($M \sim R$) with large sound-speeds ($c_s \sim 1$), it does not appear to be possible to completely restabilize the star at $p_c \geq p_{n=1}$.
Fig.~\ref{fig:compactness sequence} shows a sequence of EoSs, each of which increases $c_s^2$ within the NS core and correspondingly increases $M_\mathrm{TOV}$.
For each of these EoS, I additionally consider alternate high-$p$ extensions with $c_s^2=1$ for all $p \geq p_{n=1}$ (from the original EoS).
This shows that, although low-mass, low-$C$ stars can sometimes be restabilized in this way, high-mass, high-$C$ NSs all remain unstable even with causal extensions at $p > p_{n=1}$.

If one is additionally concerned with the stability of low-$C$ stars, it may be possible to set a termination condition at the $p_c$ where $\mathrm{n}=2$ loses stability.
In general, one could set the termination condition at the $p_c$ where higher and higher order modes lose stability and capture more and more exotic behavior.
I have not systematically explored such EoS, though, as they are unlikely to be astrophysically relevant.

\end{document}